\documentstyle[12pt,fleqn,side,aasms]{article}
\def\@doubleleading{1.5}
\def\baselinestretch{\@doubleleading}
\parskip=\baselineskip

\def\etal{{\it et al.\ }}                               

\begin{document}
\pagestyle{empty}

\title{A Search for X--ray Bright Distant Clusters of Galaxies}

\author{R. C. Nichol$\footnote{
Present Address: Department of Physics and Astronomy, University of
Chicago, 5640 S. Ellis, Chicago, Illinois 60637.}$, M. P. Ulmer}
\affil{Department of Astronomy and Astrophysics, Northwestern University, 2145
N.
Sheridan Road, Evanston, Illinois 60208.}

\author{R. G. Kron}
\affil{Fermi National Accelerator Laboratory, P. O. Box 500, Batavia, Illinois
60510.}

\author{G. D. Wirth}
\affil{Board of
Studies in Astronomy and Astrophysics, University of California, Santa Cruz,
California 95064.}

\author{D. C. Koo}
\affil{University of California Observatories, Lick Observatory and Board of
Studies in Astronomy and Astrophysics, University of California, Santa Cruz,
California 95064.}

\begin{abstract}

We present the results of a search for X--ray luminous distant clusters of
galaxies. We found extended X--ray emission characteristic of a cluster towards
two of our candidate clusters of galaxies.  They both have a luminosity in the
ROSAT bandpass of $\simeq10^{44}{\rm \,erg\,s^{-1}}$ and a redshift of $>0.5$;
thus making  them two of the most distant X--ray clusters ever observed.
Furthermore, we show that both clusters are optically rich and have a known
radio source associated  with them.  We compare our result with other recent
searches for distant X--ray luminous clusters and present a lower limit of
$1.2\times10^{-7}\,{\rm Mpc^{-3}}$ for the number density of such high redshift
clusters. This limit is consistent with the expected abundance of such
clusters in a standard (b=2)
Cold Dark Matter Universe. Finally, our clusters provide
important high redshift targets for further study into the origin and evolution
of massive clusters  of galaxies.

\end{abstract} 

\newpage
\section{Introduction}

Clusters of galaxies are the largest gravitationally bound objects in
the universe. They provide an ideal opportunity for studying the
large--scale environment of galaxies and have provided important
insights into the physics involved in the formation of both galaxies
and clusters (see Sarazin 1986). Furthermore, clusters are key tracers
of the large--scale structure in the universe, since their typical
separation is $\sim 10\,{\rm Mpc}
\!\footnote{\rm Throughout this paper, we use $H_o=50\,{\rm
km\,s^{-1}\,Mpc^{-1}}$ and $q_o=\frac{1}{2}$.} $. This represents an
efficient way of mapping the distribution of matter in the universe
({\it e.g.} Nichol {\it et al.} 1992, Guzzo {\it et al.} 1992).

The most detailed studies of clusters of galaxies, either as
individual targets or mapping their distribution, have been carried
out at relatively low redshifts ($z<0.15$). This is due to the ease
with which these objects can be observed. Clearly, it is important to
probe the universe to higher, more cosmologically interesting,
redshifts to increase the baseline over which evolutionary effects can
be studied.  The power of this approach was recently illustrated by
the claims of strong evolution in the X--ray cluster luminosity
function (Gioia \etal 1990, Henry \etal 1992).  For the largest sample
of X--ray clusters available prior to ROSAT, Henry \etal suggest that
the number density of X--ray bright clusters ($\geq10^{45}\,{\rm
erg\,s^{-1}}$) decreases rapidly as a function of lookback time, out
to a redshift of 0.6.  Some evidence for strong evolution has also
been claimed by Edge \etal (1990) but for a much nearer sample of
bright X--ray clusters ($z<0.2$). If these results are true, they
strongly suggest a hierarchical scenario of structure formation in a
high $\Omega_o$ universe (Edge \etal 1990).  Moreover, they indicate
that high redshift X--ray bright clusters are extremely rare objects
and therefore, warrant detailed study since they provide us with vital
information on the state of the universe at these large lookback
times.

In this paper, we present the results of a search for such distant
X--ray bright clusters of galaxies. The motivation behind this search
was primarily to increase number of such clusters known, thus allowing
us to form the basis for a statistical sample. In addition, they would
also provide important future targets for studying the universe at
these earlier epochs. In Section 2, we describe the methods used in
finding these clusters, while in Section 3 we detail the results of
our search.  In section 4, we discuss the consequences of our
observations and other searches for high redshift X--ray clusters.

\section{Observations and Data Analysis}

Detecting clusters at high redshift is difficult. Optically, large
areas of the sky have to be surveyed to faint limits which is
time--consuming.  Furthermore, the problems of phantom clusters, where
field galaxies and/or groups of galaxies superimposed along the
line--of--sight give the impression of a rich cluster, become severe
at these high redshifts (see Nichol 1992).  In the X--rays, we are
still in the early stages of X--ray survey astronomy with most of the
observed distant X--ray clusters gained from either specific pointing
observations, or, serendipitous discoveries.  Therefore, we started a
program to discover new distant X--ray bright clusters of galaxies
using both unidentified extended radio sources and rich optical
clusters selected from deep photographic plates as targets.

Most of the radio data used were taken from the survey of Hanish \&
Ulmer (1985), who surveyed an area of $3.4\times10^{-3}$ steradians
around nearby clusters at a frequency of 1465 MHz using the VLA.
Their original motivation was to identify wide--angled radio sources
(WAR) with bright galaxies in the outskirts of these clusters and
thus, measure the spatial extent of the hot gas. A WAR source is a
classic signature of a hot intracluster medium being present, where
the dynamical pressure from the supersonic motion of the radio galaxy
through the medium gives rise to a characteristic U--shaped bow shock
(Riley 1973).  In addition to their many optically identified radio
sources, they also discovered 17 faint resolved extended sources and 2
faint WAR sources, to a flux limit of $20\, {\rm mJys}$, which were
not identified with any object on the Palomar Observatory Sky Survey
(POSS) photographic plates.

The optical data consisted of many deep photographic
plates, taken in several optical bandpasses, on ten high--latitude
fields (see Kron 1980
\& Windhorst, Kron \& Koo 1984).
Each field has an effective radius of 25 arcminutes, thus giving an
overall survey area of $1.7\times10^{-3}$ steradians.  The ten fields
were visually scanned, on F, N and J photographic plates, individually
by R. Kron, S. Majewski and R. Dreiser and each observer compiled a
separate list of possible deep optical clusters that were seen in all
colours.  These lists were then merged and the richest clusters,
identified by all three observers, were added to our candidate cluster
list.

Table 1 lists the details of the original sample of clusters we
submitted as an AO--1 proposal for observations with ROSAT. Also
indicated in this table is the total observing time we gained with the
ROSAT PSPC for each of these cluster candidates. Four of our targets
gained enough observing time for us to search for X--ray cluster
emission.  These included one WAR source, two distant optical clusters
and a resolved extended radio source. Another WAR source gained 1300
seconds, but this was too short an integration to detect any faint
X--ray emission.

The X--ray data was processed using version 2.1 of PROS on a local
Sparc2 workstation.  The program DETECT was used to create a list of
sources for each separate pointing.  This was achieved using 3
different cellsizes; 30, 60 and 120 arcseconds, with a detection
threshold of $3\sigma$.  These were then merged into a single list of
objects and in the case of multiple entries, the detection in the
largest cellsize was used for that source.  Our next major goal was to
assess the extent of these sources since this is the strongest
indication of cluster X--ray emission.  This was not a trivial task
since the Point--Spread Function (PSF) of the ROSAT PSPC increased
dramatically with off--axis distance.  This was especially the case
for pointing number one (Table 1), which contained two of our targets
both 6 arcminutes off axis.

We therefore, determined the width of the PSF for various off-axis
distances using all X--ray sources positively identified in all our
pointings with bright stars in the SIMBAD database (7 stars) and
calibration data provided by the ROSAT team for the stars AR LAC and
HZ 43; observed at several different positions within the PSPC (13
positions). We fit the radial profiles for these 20 stars with a
gaussian, which was the observed dominant component of the on--axis
PSF gained from prelaunch tests (Hasinger \etal 1992).  We assumed
that the PSF was symmetrical and circular which from visual
inspections of our data, was a fair representation out to large
off-axis angles ($\simeq40$ arcminutes). A plot of the width of the
fitted gaussian to these stars against off--axis angle is shown in
Figure 1. Also shown is a model fit to this data ($\chi^2=14$ for 17
degrees of freedom) of the form;

\begin{equation}
\sigma(PSF)=(0.24\pm0.01) + (3.0\pm0.9\times10^{-4})\,R^{+2.35\pm0.08},
\end{equation}

\noindent where $\sigma(PSF)$ is the radius of the gaussian fitted and $R$ is
the radial angle from the centre of the PSPC (both in arcminutes).
The zero--point of this relationship agrees well with the on--axis PSF
quoted by Hasinger \etal (1992) of $\simeq 0.2$ arcminutes over the
energy passband of ROSAT.  This model therefore, provided us with a
means of determining the extent of any object detected in our PSPC
data.  For each pointing, we plotted the radial fits for all detected
objects and compared them with this relationship. We defined all
sources that were $>3\sigma$ above this model as extended.

After we had determined the extent of the detected sources, their
total counts were measured using an aperture on the source and either
an annulus around this, or, a nearby area devoid of sources to
determine the background.  The angular size of these apertures and
annulli was determined empirically and varied in size depending on the
proximity of other sources.  In almost all cases, the apertures were
much larger than the observed angular extent of the sources and
therefore, it is fair to assume that our measured source counts are
close to the total X--ray flux from the object in the ROSAT bandpass
(see below).

The measured background--corrected counts were converted to count
rates using the exposure maps provided with each pointing from the
ROSAT SASS software. For each source, the average exposure time within
the same aperture described above was calculated and used instead of
the total observing time given for the whole field.  Finally, the
hardness ratio ($HR$) for all detected sources was derived using the
same apertures. The hardness ratio was defined as; ${\rm
HR=(B-A)/(B+A)}$ where B was the background subtracted counts in the
range $0.07\rightarrow0.4{\,\rm keV}$ and A was background subtracted
counts in the range $0.4\rightarrow2.5{\,\rm keV}$. This is the same
definition as used by the standard ROSAT SASS software.

As stated above, we are only interested in extended X--ray emission,
since this is characteristic of a cluster of galaxies. The results of
our search for extended emission are summarised in Figure 1; where we
find 7 of our detected sources are extended compared to Equation 1.
These sources include two of our distant cluster candidates (marked A
and B), three known Abell clusters (A348, A350, A351 :Abell 1958) and
two large nearby galaxies seen in the 1719+49 field. The X--ray
details of the two detected distant cluster candidates are presented
in Table 2 and are discussed in detail below, while the other extended
sources will be presented in a subsequent paper.

\section{Results}

We have positively detected extended X--ray emission towards two of
our distant cluster candidates and the X--ray details of these two
clusters are presented in Table 2.  The observed hardness ratios of
our clusters can be compared to the hardness ratios of other detected
objects in our ROSAT pointings. For example, they differ significantly
($>4\sigma$) from those of identified stars and the two nearby
galaxies with extended emission mentioned above. Furthermore, the
hardness ratios of our candidate clusters are consistent with the
observed ratio of A348 ($z=0.274$, $HR=0.59$), which is detected at
high signal--to--noise in our data. Finally, the hardness ratios of
our candidates are inconsistent with the X--ray colours quoted by Kim,
Fabbiano
\& Trinchieri (1992) for active galactic nuclei.

The optical data we have obtained shows the existence of rich optical
clusters of galaxies coincident with the X--ray emission.  The
probability of a chance coincidence of an extended X--ray source with
our cluster candidates can be derived from the surface density of
extended sources in our fields. Using the central 40 arcminute region
of the PSPC, the probability of an extended X--ray source, by chance,
being within 30 arcseconds of the optical cluster is $\simeq10^{-4}$.
This distance corresponds to the largest observed uncertainty on the
positions of both the optical and X--ray centroids.  Therefore, these
observations combined strongly suggest that we have detected extended
X--ray cluster emission.  We will discuss the two detected clusters
individually below.  We note here that we did not detect any
statistically significant X--ray emission, extended or not, towards
the remaining observed cluster candidates; although we did reach a
flux sensitivity level of $1\times10^{-14}\,{\rm
ergs\,\,cm^{-2}\,s^{-1}}$ in the hard ROSAT energy passband for
pointing number four. We will no longer discuss this candidate since
only an upper limit on the X--ray flux can be obtained.

\subsection{Cl0223--09}

The centroid of the observed extended X--ray emission is only 8
arcseconds away from the WAR source presented in Table 1 and well
within the boundary of this emission$\!\footnote{\rm In the pointing
mode, the aspect solution of the PSPC is specified to be better than 6
arcseconds, with a systematic error of 2.5 arcseconds. For the weakest
sources, a statistical error of 7 arcseconds is expected (UK ROSAT
Announcement of Opportunity 1991).}$.  To verify the existence of this
cluster, we obtained a deep large--format CCD frame of this area in
December 1992 with the Isaac Newton Telescope (the observations were
kindly carried out by Stuart Lumsden and Tom Broadhurst).  A rich
cluster of galaxies is evident with its optical centroid
($\alpha=02^{h}\,23^{m}\,32.1^{s}\,\,\,\delta=08^{\circ}\,57^{'}\,48^{''}$
J2000 with an optical astrometrical uncertainty of 3 arcseconds)
within $13$ arcseconds of the X--ray centroid given in Table 2.
Furthermore, the WAR source quoted in Table 1 is only 4 arcseconds
(again with a 3 arcsecond astrometrical error) away from the brightest
optical galaxy seen within the cluster region (Figure 2). If we assume
this galaxy is a cluster member, then this observation justifies our
original motivation for observing unidentified WAR sources as
candidate distant clusters. The optical background--corrected richness
of the cluster is 51 galaxies to the limit of the CCD frame
($R\simeq21.5$) and within a 3 arcminute radius on the optical
centroid. A more robust richness estimate of Cl0223-09 will require
deeper optical observations.

We have no measured redshift for this cluster. However, using the
imaging data and the fact that it is not seen on the POSS plates, we
can estimate its redshift assuming little or no evolution in the
cluster optical luminosity function (Scaramella \etal 1991).  First,
using the universal cluster luminosity function of Colless (1989) for
rich optical clusters and redshifting it, with suitable K--corrections
(Shanks \etal 1984), the cluster must be at a $z>0.5$ to shift even
the brightest cluster member below the plate limit ($R\simeq20$).
Secondly, from the CCD image of the cluster, we have estimated the
magnitude of the tenth brightest member ($m_{10}$) to be
$R\simeq21.5$, which corresponds to an estimated redshift of $0.68$;
using the best available determination of the $m_{10}-z$ relationship
(see Nichol 1992).

Using the estimated redshift above (z=0.68), the cluster has an
estimated X--ray luminosity (k--corrected) in athe full passband of
ROSAT ($0.07\rightarrow2.5\,{\rm keV}$) of $2.3\times10^{44}\,{\rm
erg\, s^{-1}}$ within an angular aperture of radius 4 arcminutes.  As
mentioned above, this is significantly larger than the observed
angular extent of the cluster and corresponds to a spatial diameter of
$3.8$ Mpc at the estimated redshift of the cluster. This would
therefore, give us nearly 90\% of the cluster luminosity assuming a
core radius of $0.3$ Mpc and a King model with a $\beta=2/3$.  We make
no correction to the observed luminosity for the size of our aperture
since the correction factor is much smaller than overall uncertainty
in the cluster redshift and gas temperature (see below).

The luminosity derived above was obtained using FIT and XFLUX within
PROS using the rebinning facility to account for the small number of
counts in our objects. The advantages of this procedure are that, for
any given X--ray spectrum, it directly corrects for the K--correction,
photoelectric absorption by the interstellar medium and the response
function of the satellite.  However, it does require the user to
specify the form of the spectrum for the cluster. In the above
calculation, the spectrum, in the rest frame of the source, was
assumed to be a Raymond--Smith (RS) thermal plasma with a temperature
of $7\,{\rm keV}$ and $25\%$ cosmic abundance of heavy elements.  We
re--calculated its luminosity for a number of different combinations
of redshift ($z=0.4\rightarrow0.7$), model spectra ({\it i.e.}
Bremsstrahlung), temperatures ($4$ to $13$ keV) and abundance of heavy
elements.  The range in luminosities we derived from these
combinations was $9.0\times10^{43}\,{\rm erg\,s^{-1}}$ to
$3.0\times10^{44}\,{\rm erg\,s^{-1}}$ in the ROSAT bandpass.  We have
not quoted statistical errors on this result simply because the
potential systematic changes due to the unknown redshift and
properties of our distant cluster are much greater.

\subsection{Cl1719+49}

Figure 3 is a photograph of the original cluster visually found by
Kron and his collaborators on their KPNO prime--focus 4--metre
photographic plates.  The centroid of the X--ray emission is shown and
is $24$ arcseconds away from the optical centroid given in Table 1.
The radio source 53W062, published by Windhorst, Kron \& Koo (1984),
is also shown in Figure 3 and is $13$ and $34$ arcseconds away from
the optical and X--ray centroid respectively; as well as appearing to
be coincident with a galaxy within the cluster.

This cluster has a measured redshift of $z=0.61$, based on unpublished
Cryogenic Camera observations of three galaxies that are likely to be
cluster members, plus the published radio source 53W062 (Windhorst,
Kron \& Koo 1984).  Hamilton (1985) observed another galaxy at the
same redshift within the same region.  Using a RS model as above for
the X--ray spectrum, this cluster has an X--ray luminosity of
$1.4\times10^{44}\,{\rm erg\,s^{-1}}$ in the ROSAT passband and within
an angular aperture of radius 3.5 arcminutes.  The luminosity of this
cluster changed by less than $5\%$ for different combinations of model
spectra and temperatures as discussed above for Cl0223-09. No
correction has been made for the finite size of the detection aperture
since it is once again large compared to the observed extent of the
cluster.

This cluster has a background--corrected count of 32 galaxies within
the magnitude range $19<I<21$ and a radius of 3 arcminutes. This
corresponds to an absolute magnitude range of $M^{\star}-2$ to
$M^{\star}$ and a projected area of $6\,{\rm Mpc^2}$ around the
cluster centroid. In comparison, Coma has only 8 cluster members
within this magnitude range and area (Thompson
\& Gregory 1980) thus suggesting, if we assume no significant evolution of the
optical cluster luminosity function, that Cl1719+49 is a very rich optical
cluster.

\section{Discussion}

The original targets shown in Table 1 represent the most likely
candidates, based either on their galaxy richness or the presence of a
WAR source, for being the most luminous X--ray clusters in our optical
and radio survey regions.  Despite this, the X--ray luminosities of
our two clusters are consistent with the mean observed luminosity of
nearby X--ray clusters (Kowalski \etal 1984).  To extend this analysis
further and place constraints on the luminosity function of high
redshift X--ray bright clusters would require detailed knowledge of
our obviously complicated selection function ({\it i.e.} what fraction
of X--ray bright clusters do not possess a WAR source and is this a
function of luminosity?).  Furthermore, there are still other X--ray
cluster candidates in our optical and radio survey regions that have
not been observed in X--rays.  This therefore, greatly hinders us from
accurately determining our observed number density of distant X--ray
clusters, however, our search for X--ray luminous distant clusters is
qualitatively consistent with the scenario advocated by Henry \etal
(1992). They suggest that the number density of bright X--ray clusters
decreases with lookback time and we have surveyed $\simeq16\, {\rm
degrees^2}$ to faint flux limits in the optical and radio passbands
but have yet to discover any distant X--ray bright clusters
($L_x\geq10^{45}\,{\rm erg\,s^{-1}}$).

We can compare our result with other recent searches for X--ray
emission from distant clusters of galaxies and combine these data to
obtain more stringent constraints. For example, Castander \etal (1993)
has published the findings of their search for X--ray luminous
clusters from the Gunn, Hoessel \& Oke (1986) sample of distant
optical clusters. They selected the six highest redshift clusters from
this database ($z\sim0.5\rightarrow0.9$) for observation with the
ROSAT PSPC. Their results are in good agreement with ours in that they
only find statistically significant X--ray emission from three of
their clusters; all of which have a X--ray luminosity of
$\simeq10^{44}\,{\rm erg\,s^{-1}}$ in the ROSAT bandpass.

Therefore, combining the Castander {\it et al.} data and ours, we have
positive detections of five clusters in the redshift shell
$z=0.6\rightarrow0.9$ with an X--ray luminosity of
$\simeq10^{44}\,{\rm erg\,s^{-1}}$ in the ROSAT bandpass.  The maximum
possible volume surveyed by the two independent searches (assuming a
flat selection function and no evolution) can be obtained by
integrating the formula;

\begin{equation}
\frac{dV}{dz} = 4\,d\Omega\left(\frac{c}{H_o}\right)^3\,\,\,\frac{
\left(z-\sqrt{1+z}+1\right)^2}{\left(1+z\right)^{\frac{7}{2}}},
\end{equation}

\noindent ($q_o=\frac{1}{2}$, Kolb \& Turner 1990) over the entire  redshift
shell ($z=0.6\rightarrow0.9$) of our searches,  with $d\Omega$ being the total
area  covered by the  two surveys ($6.2\times10^{-3}$ steradians).  This
therefore, implies an absolute lower limit of $1.2\times10^{-7}\,{\rm
Mpc^{-3}}$ to the number density of such clusters.

This lower limit provides an important normalisation for future models
of structure formation since it is a direct observation of the
underlying mass distribution at these high redshifts, which then
evolves to form the nearby structures. This is illustrated in the
simulations of Frenk \etal (1990) where the number density of clusters
in a standard Cold Dark Matter (CDM, Davis \etal 1985) universe drops
by an order of magnitude between now and a redshift of $z=0.7$. The
lower limit quoted above is consistent with the predicted abundance of
our clusters at a $z=0.7$ in this CDM model ($b=2$, Frenk \etal 1990,
Castander \etal 1993). However, as discussed above, this is the
absolute lower limit on the abundance of such clusters and if future
observations uncover yet more clusters at these high redshifts with
similar X--ray luminosities, this would provide strong evidence for a
lowering of the biasing parameter for CDM models; which has already
been advocated by other authors from observations of distant clusters
({\it i.e.} Evrard 1989).

In addition to the observed number density of distant clusters,
detailed study of such systems at high redshift is vital to our
understanding of the state of the universe at these large lookback
time.  For example, the galaxy population of Cl1719+49 appears red
similar to that observed for the famous Cl0016+16 cluster. A detailed
analysis, similar to that performed by Butcher \& Oemler (1984), puts
the blue fraction ($f_B$) of the cluster at $f_B=0.25\pm0.24$. This
large error is primarily due to the F photographic plate not being
deep enough to give a well--determined $f_B$.  However, if subsequent
photometric studies of this cluster verify our qualitative assessment
of the colour of Cl1719+49 it is important to understand such red
clusters, at high redshift, since they conflict with the scenario
advocated by Butcher \& Oemler (1984) {\it i.e.} the fraction of blue
galaxies in clusters increases substantially with lookback time.
Furthermore, it is imperative to study known X--ray distant clusters
as the relationship between the X--ray emitting gas and the
Butcher--Oemler effect still remains relatively unknown (see Wang \&
Stocke 1993). Clearly, as the number of known clusters at high
redshift increases we will be able to create statistical samples of
such objects which will then provide us with important insights into
the nature of the universe at these significant lookback times.

\section{Acknowledgements}

Bob Nichol would like to thank many people for simulating discussions
related to this paper, these include Chris Collins, Stuart Lumsden,
Richard Ellis and Alastair Edge. We are grateful to Stuart Lumsden and
Tom Broadhurst for observing Cl0223-09, Richard Dreiser and Steve
Majewski for visually scanning the deep optical plates, Dr. Hanish for
the use of his data and John Smetanka for much help with the deep
photographic data. We thank Harvey MacGillivray for providing us with
COSMOS data in the Cl0223-09 region which allowed us to calculate an
accurate position for this cluster. Finally, we appreciate the helpful
comments of an anonymous referee. This work was carried out on NASA
grant NAG5--1633.

\newpage

\section{References}

\noindent Abell, G. O. 1958, ApJS, 3, 211\\
Butcher, H. \& Oemler, A. 1984, ApJ, 285, 426\\
Castander, F. J., Ellis, R, S., Frenk, C. S., Dressler, A. \& Gunn, J. E. 1993,
Nature, submitted\\
Colless, M., 1989, MNRAS, 237, 799\\
Davis, M., Efstathiou, G., Frenk, C.S. \& White, S.D.M. 1985, ApJ, 292, 371\\
Edge, A. C., Stewart, G. C., Fabian, A. C. \& Arnaud, K, A.
1990, MNRAS, 245, 559\\
Evrard, A. E. 1989, ApJ, 341, L71\\
Frenk, C.S., White, S.D.M., Efstathiou, G. \& Davis, M.
1990, ApJ, 351, 10\\
Gioia, L., Henry, J. P., Maccacaro, T., Morris, S. L., Stocke, J.,
\& Wolter, A. 1990, ApJ, 356, L35\\
Gunn, J. E., Hoessel, J. \& Oke, J. B. 1986, ApJ, 306, 30\\
Guzzo, L., Collins, C. A., Nichol, R. C. \& Lumsden, S. L., 1992,
ApJ, 393, L5\\
Hamilton, D. 1985, ApJ, 297, 371\\
Hanish, R. J. \& Ulmer, M. P. 1985, AJ, 90, 1407\\
Hasinger, G., Jane Turner, T., George, I. M. \& Boese, G. 1992,
ROSAT Calibration Memo\\
Henry, J. P., Gioia, L., Maccacaro, T., Morris, S. L., Stocke, J. \& Wolter, A.
1992, ApJ, 386, 408\\
Kim, D. W., Fabbiano, G. \& Trinchieri, G. 1992, ApJS, 80, 32\\
Kolb E.W., Turner M.S., 1990, The Early Universe, Addison--Wesley
Publishing Company\\
Kowalski, M. P., Ulmer, M. P., Cruddace, R. G. \& Wood, K. S.
1984, ApJS, 56, 403\\
Kron, R. G. 1980, ApJS, 43, 303\\
Nichol, R. C. 1992, Ph. D. Thesis, University of Edinburgh\\
Nichol, R. C., Collins, C. A., Guzzo, L. \& Lumsden, S. L., 1992, MNRAS,
255, 21p\\
Riley, R. 1973, MNRAS, 156, 252\\
Sarazin, C. L. 1986, Rev. Mod. Phys., 58, 1\\
Scaramella, R., Zamorani, G., Vettolani, G. \& Chincarini, G.
1991, ApJ, 101, 342\\
Shanks, T., Stevenson, P. R, F., Fong, R. \& MacGillivray, H.T., 1984, MNRAS,
206, 767\\
Stark, A.A., Gammie, C.F., Wilson, R.W., Balley, J., Linke, R.A.,
Heiles, C. \& Hurwity, M. 1992, ApJS, 79, 77\\
Thompson, L. A. \& Gregory, S. A. 1980, ApJ, 242, 1\\
Wang, Q. \& Stocke, J. T. 1993, ApJ, 408, 71\\
Windhorst, R., Kron, R. G. \& Koo, D. C. 1984, A\&AS, 39, 58\\

\newpage

\section{Table Captions}

\noindent{\bf Table 1:} Original cluster candidate list submitted for ROSAT
observations. The first two columns give the original priority and
names of our targets. The names signify: WAR as a wide--angled radio
source, EXT as a resolved extended radio source and DOC as a distant
optical cluster. The coordinates are the observed centroids of these
objects. Column 5 is the total amount of ROSAT PSPC observing time
gained for that specific pointing, while Column 6 gives the radio
fluxes quoted by Hanish \& Ulmer (1985), except for DOC1 which has a
published radio source (Windhorst, Kron \& Koo 1984). Finally, the
last column indicates whether a optical cluster was known in that
direction before the ROSAT observations.

\vspace*{0.01in}

\noindent{\bf Table 2:} The two clusters positively
detected in X--rays from the list of candidates presented in Table 1.
For the remaining candidates, no X--ray source was seen within 60
arcseconds of the quoted candidate position.  The first column is the
candidate name given in Table 1.  The coordinates in the next two
columns are the X--ray centroid of the emission seen. Column 4
presents the raw net counts, and the error, measured in the energy
band $0.5\rightarrow2.4$keV.  Columns 5 and 6 are the X--ray flux and
luminosity calculated in the full ROSAT bandpass
($0.07\rightarrow2.5{\,\rm keV}$). The hardness ratios are given in
Column 7.  Column 8 gives the full observed extent of the cluster,
followed by the implied cluster extent after the contribution from the
PSPC PSF has been removed (assuming a gaussian profile for the cluster
and PSF).  The errors on these extents are also presented.  The final
two columns are the hydrogen column density from the Stark
\etal (1992) data and the specific ROSAT exposure time gained by the
object which includes the effects of vignetting.

\newpage

\section{Figure Captions}

\noindent{\bf Figure 1:} This figure shows the observed relationship between
the width of the ROSAT PSPC Point--Spread Function against off--axis
angle.  The black symbols ($\bullet$) represent known stars used to
calibrate this relationship.  Error bars are plotted for all these
points but some are smaller than the plotting symbol used.  The best
fit to this data is also shown as a dashed line (Equation 1 in the
main text).  The triangular symbols ($\triangle$) are extended sources
detected within all our PSPC pointings, while the diamond symbols
($\Diamond$) are the two clusters reported in this paper. Point A is
Cl1719+49 and point B is Cl0223-09.

\vspace*{0.02in}

\noindent{\bf Figure 2:} This is a deep CCD image taken in December 1992 of
cluster Cl0223-09. The size of the image shown is 3 arcminutes square
and reaches a limiting magnitude of $R\simeq22.5$. A rich cluster of
galaxies is evident near both the X--ray centroid ($+$) and the
wide--angled radio source ($\times$).  The bar shown represents 30
arcseconds.

\vspace*{0.02in}

\noindent{\bf Figure 3:} This is a photograph of the distant optical cluster
originally discovered by Richard Kron and his collaborators.  The
whole region shown is 11 by 7 arcminutes taken from the N--band
photographic plate. The X--ray centroid ($\times$) and radio source
53W062 ($\Box$) are marked and are clearly coincident with the rich
optical cluster.

\newpage

\begin{table}[htbf]
\begin{tabular}{c|cccccc}
Pointing & Name     &  RA    & DEC    & ROSAT Total & Radio Flux &Optical\\
No.      &          &(J2000) & (J2000)&  Obs. Time  &  (mJys)    &Cluster\\
\hline
1        & WAR1     &02 23 32.7 & $-08$ 57 43 & 13797  &93        & no \\
1        & EXT1     &02 23 10.0 & $-09$ 08 45 & 13797  &39        & no \\
2        & DOC1     &17 19 31.0 & $+49$ 59 06 & 33579  &$2^{\star}$ & yes \\
3        & WAR2     &03 03 31.7 & $+41$ 22 33 & 1300   &88        & no  \\
4        & DOC2     &08 49 15.1 & $+44$ 28 37 & 15703  &            & yes \\
5        & EXT2     &12 38 46.1 & $+03$ 23 20 & 0      &32          & no  \\
\end{tabular}
\end{table}

\newpage

{\small
\begin{sidetable}
\begin{tabular}{c|cccccccccc}
Name     &  RA    & DEC    & Raw Net & Flux & Lum& Hardness& Obs. Extent&
Corr. Extent&${\rm log_{10}(N_H)}$& Exp\\
         &(J2000) & (J2000)& Counts  &(${\rm erg\, cm^{-2}
\,s^{-1}}$)    &(${\rm erg\, s^{-1}})$&Ratio& ('')& ('') &(atoms ${\rm
cm^{-2}}$) &(secs)\\ \hline
 WAR1     &02 23 33.0 & $-08$ 57 50 & $94\pm13$ &$9.2\times10^{-14}$
&$2.3\times10^{44}$ & $0.95\pm0.26$&$62\pm7$&$57\pm 9$ & 20.49 &13499\\
 DOC1     &17 19 29.6 & $+49$ 59 18 &  $139\pm16$&$7.3\times10^{-14}$
&$1.4\times10^{44}$ & $0.64\pm0.10$&$48\pm6$&$45\pm 8$ & 20.37 &30792\\
\end{tabular}
\end{sidetable}
}

\end{document}